\documentclass[11pt]{amsart}
\usepackage[english]{babel}
\usepackage{egothic}
\usepackage[T1]{fontenc}

\usepackage{amsmath}
\usepackage{amsthm}
\usepackage{amssymb}
\usepackage[all]{xy}
\usepackage{mathrsfs}
\usepackage{enumerate}
\usepackage{physics}
\usepackage[notcite, final, notref]{showkeys}
\usepackage{dsfont}
\usepackage[dvips]{color}
\newtheorem{theorem}{Theorem}[section]

\newtheorem{lemma}[theorem]{Lemma}
\newtheorem{proposition}[theorem]{Proposition}
\newtheorem{corollary}[theorem]{Corollary}
\newtheorem{definition}[theorem]{Definition}

\theoremstyle{definition}
\newtheorem{remark}[theorem]{Remark}

\newtheorem{example}[theorem]{Example}

\usepackage{xcolor}

\title[]{Superoscillations and Fock spaces}

\author[D. Alpay]{Daniel Alpay}
\address{(DA) Schmid College of Science and Technology \\
Chapman University\\
One University Drive
Orange, California 92866\\
USA}
\email{alpay@chapman.edu}

\author[F. Colombo]{Fabrizio Colombo}
\address{(FC) Politecnico di
Milano\\Dipartimento di Matematica\\Via E. Bonardi, 9\\20133 Milano\\Italy}
\email{fabrizio.colombo@polimi.it}

\author[K. Diki]{Kamal Diki}
\address{(KD) Schmid College of Science and Technology \\
Chapman University\\
One University Drive
Orange, California 92866\\
USA}
\email{diki@chapman.edu}

\author[I. Sabadini]{Irene Sabadini}
\address{(IS) Politecnico di
Milano\\Dipartimento di Matematica\\Via E. Bonardi, 9\\20133 Milano\\Italy}
\email{irene.sabadini@polimi.it}

\author[D. C. Struppa]{Daniele C. Struppa}
\address{(DCS) Schmid College of Science and Technology \\
Chapman University\\
One University Drive
Orange, California 92866\\
USA}
\email{struppa@chapman.edu}

\begin{document}
\maketitle
\begin{abstract}
In this paper we use techniques in Fock spaces theory and compute how the Segal-Bargmann transform acts on special wave functions obtained by multiplying superoscillating sequences with normalized Hermite functions. It turns out that these special wave functions can be constructed also by computing the approximating sequence of the normalized Hermite functions. First, we start by treating the case when a superoscillating sequence is multiplied by the Gaussian function. Then, we extend these calculations to the case of normalized Hermite functions leading to interesting relations with Weyl operators. In particular, we show that the Segal-Bargmann transform maps superoscillating sequences onto a superposition of coherent states. Following this approach, the computations lead to a specific linear combination of the normalized reproducing kernels (coherent states) of the Fock space. As a consequence, we obtain two new integral Bargmann-type representations of superoscillating sequences. We also investigate some results relating superoscillation functions with Weyl operators and Fourier transform.
\end{abstract}



\tableofcontents

\section{Introduction}
\setcounter{equation}{0}

In 1961 Bargmann introduced in \cite{Bargmann1961} a Hilbert space of entire functions in which the creation and annihilation operators, namely $$\displaystyle M_zf(z):=zf(z)\quad \text{and} \quad Df(z):=\frac{d}{dz}f(z)$$ are closed, densely defined operators, adjoints of each other and satisfy the classical commutation rule 
$$\left[D,M_z\right]=\mathcal{I}$$ 

where $\left[.,.\right]$ and $\mathcal{I}$ are the commutator and the identity operator respectively. This space is called Fock (or Segal-Bargmann) space. It consists of entire functions that are square integrable on the complex plane with respect to the normalized Gaussian measure. The Fock space is a fundamental mathematical model in quantum mechanics since it is unitary equivalent to the standard $L^2$-Schr\"odinger Hilbert space on the real line via the so-called Segal-Bargmann transform discovered in \cite{Bargmann1961}. Basically, the Segal-Bargmann transform provides a new perspective on quantum mechanics by expressing wave functions as entire, complex-valued functions.  It turns out that the Segal-Bargmann transform has many applications in different areas of physics, including path integrals, coherent states, and quantum field theory, see \cite{qtm-book}. For a complete description on Fock spaces, Segal-Bargmann theory and their applications in mathematical physics we refer the reader to the books \cite{qtm-book, neretin, zhu}. It is worth mentioning that in the last few years several authors studied Fock spaces from different perspectives and in various settings. For recent results and extensions on these spaces in complex and hypercomplex analysis, see \cite{AlpayColomboSabadini2014, ACK, BESC2019, CSS2016, CD2017, KA12, KA, DG2017, DKS2019, KMNQ2016, PSS2016, mlf}.
\\ \\
Over the last five decades, quantum physicists proved the existence of a phenomenon known as \textit{superoscillation}. This interesting concept was found to occur naturally when working with weak measurements in quantum mechanics. Mathematically, superoscillations refer to the combination of small Fourier components, whose spectrum is bounded, resulting in a shift that can be much larger than the spectrum itself, when suitably added together. The work of Aharonov and his collaborators has shed light on this fascinating phenomenon and its potential applications, see \cite{ACSSTbook2017, ACSST2011JPA, BZACSSTRQHL2019} and the references therein. The main purpose of this paper is to start investigating new research results relating the theory of Fock spaces with the theory of superoscillations.
\\ \\ 
The plan of the paper is the following: in Section 2 we collect some preliminary results on Fock spaces and superoscillations. In Section 3 we show how the Segal-Bargmann transform acts on special wave functions involving superoscillation sequences. We study three different cases including the Gaussian function, Hermite functions and general $L^2$-functions. As a consequence, we prove a first integral representation of the superoscillation sequences. Section 4 is devoted to study the entire superoscillation sequence and its connection with the inverse Segal-Bargmann transform. This leads to a second integral representation of superoscillating functions. In Section 5 we obtain new results relating Fourier transform, approximating sequences and Weyl operators on Fock spaces.
\section{Preliminary results}
\setcounter{equation}{0}
In this section we recall some basic notions including Fock spaces, Segal-Bargmann transform, Hermite functions, Weyl operators, and superoscillations. For more details on these topics we refer the reader to the books \cite{ACSSTbook2017, neretin, zhu} and the references therein. \\ \\
The Fock space is a subspace of the space $H(\mathbb{C})$ of entire functions, here denoted by $\mathtt{F}(\mathbb{C})$ and defined by
$$\displaystyle \mathtt{F}(\mathbb{C}):=\lbrace{f\in H(\mathbb{C}), \quad ||f||_{\mathtt{F}(\mathbb{C})}^2=\frac{1}{\pi}\int_{\mathbb{C}}|f(z)|^2e^{-|z|^2}d\lambda(z)<+\infty}\rbrace;$$
where $d\lambda(z)=dxdy$ denotes the classical Lebesgue measure with $z=x+iy$.

The reproducing kernel of $\mathtt{F}(\mathbb{C})$ is the kernel function given by
\begin{equation}
K(z,w)=K_w(z):=e^{z\overline{w}}, \quad \forall z,w\in \mathbb{C}.
\end{equation}
The normalized reproducing kernel of the Fock space is
\begin{equation}
 k(z,w)=k_w(z)=\frac{K(z,w)}{||K_w||_{\mathtt{F}(\mathbb{C})}}=e^{z\overline{w}-\frac{|w|^2}{2}}, \quad \forall z,w \in\mathbb{C}.
\end{equation}

\begin{remark}
The reproducing kernel property on the Fock space $\mathtt{F}(\mathbb{C})$ can be expressed in terms of the following integral representation: For every $f\in\mathtt{F}(\mathbb{C})$ we have
\begin{equation}
f(w)=\displaystyle \int_{\mathbb{C}} f(z)\overline{K(z,w)}d\mu(z),
\end{equation}
where $d\mu(z)=\left(\dfrac{1}{\pi}\right)\exp(-|z|^2)d\lambda(z).$
\end{remark}
Let $w\in\mathbb{C}$ fixed, the normalized Fock kernel is given by the formula
\begin{equation}
k_{w}(z):=\dfrac{K(z,w)}{\sqrt{K(w,w)}}, \quad z\in\mathbb{C}.
\end{equation}
In particular, we have \begin{equation}
k_{w}(z):=\exp\left(z\overline{w}-\frac{|w|^2}{2}\right), \quad \forall z, w\in\mathbb{C}.
\end{equation}
We recall the Weyl operators on the Fock spaces (see \cite{qtm-book, zhu})
\begin{definition}[Weyl operator]
Let $a\in\mathbb{C}$. Then, the Weyl operator $\mathcal{W}_{a}:\mathtt{F}(\mathbb{C})\longrightarrow \mathtt{F}(\mathbb{C}),$ is defined as
\begin{equation}
\mathcal{W}_a f(z):=f(z-a)k_a(z), \quad f\in \mathtt{F},  z, a\in \mathbb{C}.
\end{equation}
\end{definition}

We summarize some interesting properties of Weyl operators in the following:
\begin{proposition}
Let $a, b\in \mathbb{C}$. The Weyl operator $\mathcal{W}_a$ defines an unitary operator from the Fock space $\mathtt{F}(\mathbb{C})$ onto itself and its inverse (and adjoint) is given by 

\begin{equation}
(\mathcal{W}_a)^{-1}=(\mathcal{W}_a)^{*}=\mathcal{W}_{-a}.
\end{equation}
Moreover, the following property holds
\begin{equation}\label{Weylsg}
\mathcal{W}_a\mathcal{W}_b=\exp(- i  \rm Im(a\bar{b}))\mathcal{W}_{a+b}, \quad a,b\in\mathbb{C}.
 \end{equation}
\end{proposition}
\begin{remark}
If $a,b\in\mathbb{R}$, the Weyl operator will satisfy the classical semi-group property
\begin{equation}\label{Weylsg1}
\mathcal{W}_a\mathcal{W}_b=\mathcal{W}_{a+b}.
 \end{equation}
\end{remark}

 Another fundamental tool in the theory of Fock spaces is the so-called Segal-Bargmann transform which makes the standard Schrödinger Hilbert space $L^2(\mathbb{R})$ unitary isomorphic to the Fock space $\mathtt{F}(\mathbb{C})$. This integral transform was introduced in \cite{Bargmann1961} by Bargmann as follows: given a function $\varphi\in L^2(\mathbb{R})$ its Segal-Bargmann transform is
\begin{equation}\label{SBDEF}
 \displaystyle \mathcal{B}(\varphi)(z):=\left(\frac{1}{\pi}\right)^{1/4}\int_\mathbb{R}e^{-\frac{1}{2}(z^2+x^2)+\sqrt{2}zx}\varphi(x)dx, \quad \forall z\in \mathbb{C}.
\end{equation}
It is possible to introduce this integral transform using the Bargmann kernel $A:\mathbb{C}\times \mathbb{R}\longrightarrow \mathbb{C}$ which is defined by 
\begin{equation}
A(z,x)=A_z(x)=A_x(z)= \left(\frac{1}{\pi}\right)^{1/4}e^{-\frac{1}{2}(\overline{z}^2+x^2)+\sqrt{2}\overline{z}x}, \quad \forall (z,x)\in \mathbb{C}\times \mathbb{R}.
\end{equation}
Then, the Segal-Bargmann transform is
\begin{equation}
\mathcal{B}(\varphi)(z):=\langle \varphi, A_z \rangle_{L^2(\mathbb{R})}, \quad \forall z\in \mathbb{C}.
\end{equation}

It turns out that the Fock kernel $K(z,w)$ can be expressed as an inner product of Bargmann kernels on the Hilbert space $L^2(\mathbb{R})$. In fact, the following expression holds true 

\begin{equation}
\langle A_w, A_z \rangle_{L_2(\mathbb{R})}=e^{z\overline{w}}=K(z,w),
\end{equation}
for every $z,w\in \mathbb{C}$. Moreover, it is well-known that the Segal-Bargmann transform $\mathcal{B}$ is an unitary integral transform mapping $L^2(\mathbb{R})$ onto the Fock space $\mathtt{F}(\mathbb{C})$ with an inverse transform given by 
 
 \begin{equation}
 \displaystyle \mathcal{B}^{-1}(f)(x)= \langle f, \overline{A_{x}} \rangle_{\mathtt{F}(\mathbb{C})}=\left(\frac{1}{\pi}\right)^{1/4}\int_\mathbb{C}e^{-\frac{1}{2}(\bar{z}^2+x^2)+\sqrt{2}\bar{z}x}f(z)e^{-|z|^2}d\lambda(z), \quad f\in\mathtt{F}(\mathbb{C}), x\in \mathbb{R}.
 \end{equation}

It is useful to recall the Rodrigues formula allowing to define Hermite polynomials:
\begin{equation}
\displaystyle H_k(x)=(-1)^ke^{x^2}\frac{\partial^k}{\partial x^k}\left(e^{-x^2}\right), \quad k=0,1,\ldots
\end{equation}
The normalized Hermite functions $h_n(x)$ are obtained by multiplying the classical Hermite polynomials $H_n(x)$ with a suitable Gaussian function and normalizing with respect to $L^2$-norm. It turns out that the Segal-Bargmann transform maps the normalized Hermite functions onto an orthonormal basis of the Fock space so that for every $k=0,1, \ldots$ we have 
\begin{equation}
\displaystyle \mathcal{B}(h_k)(z)=\frac{z^k}{\sqrt{k!}}:=e_k(z), \quad \forall z\in \mathbb{C}.
\end{equation}
  \\ \\
Now let us briefly review some basic notions related to the mathematics of superoscillations, inspired from the book \cite{ACSSTbook2017}. The prototypical superoscillating function, that appears in the theory of weak values,  is
\begin{equation}\label{FNEXP}
F_n(x,a)
=\sum_{j=0}^nC_j(n,a)e^{i(1-\frac{2j}{n})x},\ \ x\in \mathbb{R},
\end{equation}
where $a>1$ and the coefficients $C_j(n,a)$ are given by
\begin{equation}\label{Ckna}
C_j(n,a)=\binom{n}{j}\left(\frac{1+a}{2}\right)^{n-j}\left(\frac{1-a}{2}\right)^j.
\end{equation}
If we fix $x \in \mathbb{R}$ we  obtain that
$$
\lim_{n \to \infty} F_n(x,a)=e^{iax},
$$
and the limit is uniform on compact subsets of the real line. The term superoscillations comes from the fact that
in the Fourier representation of the function \eqref{FNEXP} the frequencies
$1-2j/n$ are bounded by 1, but the limit function $e^{iax}$ has a frequency $a$ that can be arbitrarily larger  than $1$.

Inspired by this example we define a {\em generalized Fourier sequence}.
These are sequences of the form
\begin{equation}\label{basic_sequenceq}
f_n(x):= \sum_{j=0}^n Z_j(n,a)e^{ih_j(n)x},\ \ \ n\in \mathbb{N},\ \ \ x\in \mathbb{R},
\end{equation}
where $a\in\mathbb R$, $Z_j(n,a)$ and $h_j(n)$
are complex and real valued functions of the variables $n,a$ and $n$, respectively.
The sequence \eqref{basic_sequenceq}
 is said to be {\em a superoscillating sequence} if
 $\displaystyle\sup_{j,n}|h_j(n)|\leq 1$ and
 there exists a compact subset of $\mathbb R$,
 which will be called {\em a superoscillation set},
 on which $f_n(x)$ converges uniformly to $e^{ig(a)x}$,
 where $g$ is a continuous real valued function such that $|g(a)|>1$.

 The notion of superoscillation can be generalized to the case in which the limit function is not an exponential. In this case we have the notion of supershift:

\begin{definition}[Supershift Property]\label{supershift}
 Let $\lambda\mapsto \varphi_\lambda(X)$ be a continuous complex-valued function
in the variable $\lambda \in \mathcal{I}$, where $\mathcal{I}\subseteq\mathbb{R}$ is an interval, and
$X\in \Omega $, where $\Omega$ is a domain.
We consider  $X\in \Omega $ as a parameter for the function
 $\lambda\mapsto \varphi_\lambda(X)$ where $\lambda \in \mathcal{I}$.
When  $[-1,1]$ is contained into $\mathcal{I}$ and $a\in \mathcal{I}$, we define the sequence
\begin{equation*}
\label{psisuprform}
\psi_n(X)=\sum_{j=0}^nC_j(n,a)\varphi_{1-\frac{2j}{n}}(X),
\end{equation*}
in which $\varphi_{\lambda}$ is computed just at the points $1-\frac{2j}{n}$ which belong to the interval $[-1,1]$ and the coefficients
$C_j(n,a)$ are defined for example as in \eqref{Ckna}, for $j=0,...,n$ and $n\in \mathbb{N}$.
If
$$
\lim_{n\to\infty}\psi_n(X)=\varphi_{a}(X)
$$
for $|a|>1$ arbitrary large (but belonging to $\mathcal{I}$), we say that the function
$\lambda\mapsto \varphi_{\lambda}(X)$,  for $X$ fixed, admits the supershift property.
\end{definition}

 \medskip
 If we set $\varphi_\lambda(x)=e^{i\lambda x}$ and $X=x\in \mathbb{R}$, we obtain the superoscillating sequence described above as a particular case of the
supershift. In fact,
in this case, we have $\psi_n(x)= F_n(x,a)$, where $F_n(x,a)$ is defined in \eqref{FNEXP}.
The name supershift is due to the fact that we are able to obtain $\varphi_{a},$ for $|a|>1$ arbitrarily large, by simply calculating
 the function
$\lambda\mapsto \varphi_{\lambda}$ at infinitely many points in the neighborhood $[-1,1]$ of the origin.

%Finally, we briefly recall the classical superoscillatory sequence of a real variable. Indeed, let $a>1$ be a real number, the classical superoscillation sequence is defined by \begin{equation}
%\displaystyle F_n(x,a):=\left(\cos\left(\frac{x}{n}\right)+i a\sin\left(\frac{x}{n}\right) \right)^n=\sum_{j=0}^{n}C_j(n,a)e^{i(1-2j/n)x}
%\end{equation}
%where $$C_j(n,a)={n\choose j} \left(\frac{1+a}{2}\right)^{n-j}\left(\frac{1-a}{2}\right)^j.$$ It is a well-known that if $x\in\mathbb{R}$ is fixed and $n$ go to infinity, we have \begin{equation}
%\lim_{n\rightarrow +\infty}F_n(x,a)=e^{iax}.
%\end{equation}
Given $a\in \mathbb{C}$ we can define the complex superoscillating sequence by
\begin{equation}\label{CompSuper}
\displaystyle F_n(z,a):=\left(\cos\left(\frac{z}{n}\right)+i a\sin\left(\frac{z}{n}\right) \right)^n.
\end{equation}
The following interesting fact holds:
\begin{lemma}
For any $a\in\mathbb{C}$, the sequence $(F_n(z,a))_{n\geq 1}$ converges to the function $F(z,a)=e^{iaz}$ in $A_1(\mathbb{C})$.
\end{lemma}

\section{Action of Segal-Bargmann transform on super-oscillations}
\setcounter{equation}{0}
In this section we compute how the Segal-Bargmann transform acts on superoscillations. We first treat the Gaussian case and then move to the case involving the normalized Hermite functions.
\subsection{First case: Gaussian function}
Inspired from the relation between Hermite polynomials and Hermite functions, we introduce a class of wave functions $\tilde{F}_n(x,a)$ obtained by  multiplying the classical superoscillating sequences $F_n(x,a)$ by the Gaussian function $e^{-x^2/2}$.

We introduce the modified superoscillating functions given by:
 \begin{definition}
 For $a>1$, we define
 \begin{equation}
 \displaystyle \tilde{F_n}(x,a):=e^{-x^2/2}F_n(x,a), \quad  x\in \mathbb{R}.
 \end{equation}
 \end{definition}
 \begin{proposition}
 For every $a>1$, it holds that
 \begin{equation}
\tilde{F}(x,a):=\lim_{n\rightarrow+\infty}\tilde{F_n}(x,a)=e^{-x^2/2+iax}, \quad x\in \mathbb{R}.
 \end{equation}
 \end{proposition}
 \begin{proof}
 This fact is based on easy calculations using that $$\lim_{n\rightarrow+\infty}F_n(x,a)=e^{iax}.$$
 \end{proof}
 \begin{remark}
  The function $\tilde{F_n}(x,a)$ belongs to the space $L^2(\mathbb{R})$. Therefore, we can compute its associated Segal-Bargmann transform leading to a special sequence of entire functions belonging to the Fock space $\mathtt{F}(\mathbb{C})$.
 \end{remark}
 We recall an interesting technical lemma which will be useful in the sequel for some technical computations:
 \begin{lemma}[Gaussian integral]\label{GAUSSIANINT}
 Let $\alpha>0$ and $w\in\mathbb{C}$. It holds that
 \begin{equation}
 \displaystyle \int_{\mathbb{R}}e^{-\alpha t^2+wt}dt=\left(\frac{\pi}{\alpha}\right)^{1/2}e^{\frac{w^2}{4\alpha}}.
 \end{equation}
 \end{lemma}
 In particular, the action of Segal-Bargmann transform on $\tilde{F}_n(x,a)$ is given by the following result:
 \begin{theorem} \label{BSO}
 Let $a>1$, for any $z\in \mathbb{C}$ we have
 \begin{equation}
\displaystyle \mathcal{B}(\tilde{F_n})(z)=\pi^{1/4}\sum_{j=0}^{n}C_j(n,a)e^{\frac{iz}{\sqrt{2}}(1-2j/n)-\frac{1}{4}(1-2j/n)^2}.
 \end{equation}

 \end{theorem}
\begin{proof}
We note that $\tilde{F_n}$ belongs to $L^2(\mathbb{R})$. Therefore, we can compute its Bargmann transform using \eqref{SBDEF}:  \begin{align*}
\displaystyle \mathcal{B}(\tilde{F_n})(z)&=\left(\frac{1}{\pi}\right)^{1/4}\int_\mathbb{R}e^{-\frac{1}{2}(z^2+x^2)+\sqrt{2}zx}\tilde{F_n}(x,a)dx\\
&=\left(\frac{1}{\pi}\right)^{1/4}\int_\mathbb{R}e^{-\frac{1}{2}(z^2+x^2)+\sqrt{2}zx}e^{-\frac{x^2}{2}}F_n(x,a)dx \\
&=\left(\frac{1}{\pi}\right)^{1/4}\sum_{j=0}^{n}C_j(n,a)\left(\int_\mathbb{R}e^{-\frac{z^2}{2}-x^2+\sqrt{2}zx} e^{i(1-2j/n)x}dx \right)\\
&=\left(\frac{1}{\pi}\right)^{1/4}e^{-\frac{z^2}{2}}\sum_{j=0}^{n}C_j(n,a)\left(\int_\mathbb{R}e^{-x^2+x(\sqrt{2}z+i(1-2j/n))} dx \right).\\
\end{align*}
Now, we use the Gaussian integral given by Lemma \ref{GAUSSIANINT} with parameters $\alpha=1$ and $w=(\sqrt{2}z+i(1-2j/n))$. So, for every $ j=0,\cdots, n$ we obtain
\begin{align*}
\displaystyle \int_\mathbb{R}e^{-x^2+x(\sqrt{2}z+i(1-2j/n))} dx&=\sqrt{\pi}e^{\frac{\left(\sqrt{2}z+i(1-2j/n)\right)^2}{4}}\\
&=\sqrt{\pi}e^{\frac{z^2}{2} +\frac{iz}{\sqrt{2}}(1-2j/n)-\frac{1}{4}(1-2j/n)^2}.\\
\end{align*}
Hence, inserting this expression in the previous computations we obtain
$$ \displaystyle \mathcal{B}(\tilde{F_n})(z)=\pi^{1/4}\sum_{j=0}^{n}C_j(n,a)e^{\frac{iz}{\sqrt{2}}(1-2j/n)-\frac{1}{4}(1-2j/n)^2}.$$
\end{proof}

  \begin{corollary}\label{Cor1}
 Let $a>1$, for every $z\in \mathbb{C}$ we have
  \begin{equation}
 \displaystyle \mathcal{B}(\tilde{F_n})(z)=\pi^{1/4}\sum_{j=0}^{n}C_j(n,a)k_{b_j}(z),
 \end{equation}
 where $b_j=-\frac{i}{\sqrt{2}}(1-2j/n)$ and $k_{b_j}$ is the normalized reproducing kernel of the Fock space $\mathcal{F}(\mathbb{C})$.
 \end{corollary}
 \begin{proof}
 It is enough to apply Theorem \ref{BSO} and observe that

 $$k_{b_j}(z)=e^{\frac{iz}{\sqrt{2}}(1-2j/n)-\frac{1}{4}(1-2j/n)^2},$$
 for every $z\in\mathbb{C}$ where $b_j=-\frac{i}{\sqrt{2}}(1-2j/n)$ for every $j=0,\cdots, n$.
 \end{proof}

  \begin{theorem}\label{Blim}
 Let $a>1$. Then, for every $z\in \mathbb{C}$ we have
 \begin{equation}
 \displaystyle \mathcal{B}(\tilde{F})(z,a)=\pi^{1/4} e^{\frac{iza}{\sqrt{2}}-\frac{a^2}{4}}.
 \end{equation}

 \end{theorem}
 \begin{proof}
 We note that $\tilde{F}$ is in $L^2(\mathbb{R})$ so that we can compute its Bargmann transform as follows \begin{align*}
\displaystyle \mathcal{B}(\tilde{F})(z)&=\left(\frac{1}{\pi}\right)^{1/4}\int_\mathbb{R}e^{-\frac{1}{2}(z^2+x^2)+\sqrt{2}zx}\tilde{F}(x,a)dx\\
&=\left(\frac{1}{\pi}\right)^{1/4}\int_\mathbb{R}e^{-\frac{1}{2}(z^2+x^2)+\sqrt{2}zx}e^{-\frac{x^2}{2}}F(x,a)dx \\
&=\left(\frac{1}{\pi}\right)^{1/4} e^{-\frac{z^2}{2}}\int_\mathbb{R}e^{-x^2+x(\sqrt{2}z+ia)}dx \\
&=\left(\frac{1}{\pi}\right)^{1/4}\sqrt{\pi} e^{-\frac{z^2}{2}} e^{\frac{(\sqrt{2}z+ia)^2}{4}}\\
&=\pi^{1/4} e^{\frac{iza}{\sqrt{2}}-\frac{a^2}{4}}.\\
\end{align*}

 \end{proof}
  \begin{corollary}\label{Cor2}
 Let $a>1$, for every $z\in \mathbb{C}$ we have
  \begin{equation}
 \displaystyle \mathcal{B}(\tilde{F})(z,a)=\pi^{1/4}k_{-ia/\sqrt{2}}(z).
 \end{equation}
 \end{corollary}
 \begin{proof}

 It is enough to apply Theorem \ref{Blim} and observe that

 $$k_{b}(z)= e^{\frac{iza}{\sqrt{2}}-\frac{a^2}{4}},$$
 for every $z\in\mathbb{C}$ where $b=-ia/\sqrt{2}$.
 \end{proof}
 \begin{proposition}\label{LIMBFN}
 Let $a>1$ and $z\in\mathbb{C}$. Then,
 \begin{equation}
 \lim_{n\rightarrow \infty} \mathcal{B}(\tilde{F}_n)(z)=\mathcal{B}(\tilde{F})(z).
 \end{equation}
 \end{proposition}
 \begin{proof}
 It is important to note that the Bargmann transform $\mathcal{B}(\tilde{F})$ is by construction an entire function, therefore, using the supershift property and the expression obtained in Theorem \ref{BSO} we can easily conclude that
$$ \lim_{n\rightarrow \infty} \mathcal{B}(\tilde{F}_n)(z)=\mathcal{B}(\tilde{F})(z).$$ 
 %We observe that $k_b(z)$ is an entire function, thus in particular it satisfies the super-shift property. 
 %First we observe using properties of the Segal-Bargmann transform that we have
 %$$|\mathcal{B}(\tilde{F_n})(z)-\mathcal{B}(\tilde{F})(z)|=|\mathcal{B}(\tilde{F_n}-\tilde{F})(z)|\leq e^{\frac{|z|^2}{2}}||\tilde{F_n}-\tilde{F}||_{L^2(\mathbb{R})}.$$
% Since $\mathcal{B}$ is linear and isometric from $L^2(\mathbb{R})$ onto $\mathit{F}(\mathbb{C})$ we have also
 %$$||\mathcal{B}(\tilde{F_n})-\mathcal{B}(\tilde{F})||_{\mathtt{F}(\mathbb{C})}=||\mathcal{B}(\tilde{F_n}-\tilde{F})||_{\mathtt{F}(\mathbb{C})}=||\tilde{F_n}-F||_{L^2(\mathbb{R})}, \quad \forall n\geq 1.$$
 %{\color{red} APPLY THE RESULT EXPLAINED BY FABRIZIO!!!!
% We still need to check using the Lebesgue dominated convergence theorem if $||\tilde{F_n}-\tilde{F}||_{L^2(\mathbb{R})}\longrightarrow 0$ when $n$ tends to $+\infty$ ???
%}
 \end{proof}
 \begin{corollary}
 Let $a>1$ and $z\in\mathbb{C}$. Then, it holds that
  \begin{equation}
 \displaystyle \lim_{n\rightarrow \infty} \sum_{j=0}^{n}C_j(n,a)e^{\frac{iz}{\sqrt{2}}(1-2j/n)-\frac{1}{4}(1-2j/n)^2}= e^{\frac{iza}{\sqrt{2}}-\frac{a^2}{4}}.
 \end{equation}
In terms of the normalized reproducing kernels of the Fock space we have

\begin{equation}\label{li}
 \displaystyle \lim_{n\rightarrow \infty} \sum_{j=0}^{n}C_j(n,a)k_{b_j}(z)=k_{-ia/\sqrt{2}}(z),
 \end{equation}

 for every $z\in\mathbb{C}$ where $k_w$ is the normalized reproducing kernel of the Fock space $\mathtt{F}(\mathbb{C})$ with parameter $w\in\mathbb{C}$.
 \end{corollary}
 \begin{proof}
This fact can be seen as a direct consequence of Corollary \ref{Cor1}, Corollary \ref{Cor2}, and Theorem \ref{LIMBFN} .
 \end{proof}
 \begin{remark}
 We note that the normalized reproducing kernels $(k_{b_j})_{j\geq 0}$ are called coherent states in quantum mechanics. Therefore, the expression \eqref{li} can be interpreted in quantum mechanics, see \cite{AAG2000Coh, A2015}. In particular, the Segal-Bargmann transform maps superoscillations onto a finite superposition of the coherent states $(k_{b_j})_{j\geq 0}$ with parameters $(b_j)_j$. Moreover, the limit case when $n\longrightarrow\infty$ leads to the coherent state $k_{-ia/\sqrt{2}}$ with the parameter $-ia/\sqrt{2}$.
 \end{remark}
The superoscillating sequences can be obtained by taking the Bargmann inverse of a superposition of coherent states leading to the following integral representation of $F_n(x,a)$:
 \begin{proposition}[Integral representation I]
 Let $a>1$ and $n\in \mathbb{N}$. We set $\displaystyle \varphi_n(z)=\sum_{j=0}^{n}C_j(n,a)k_{b_j}(z)$ with $b_j=-\frac{i}{\sqrt{2}}(1-2j/n)$. Then 
 \begin{equation}
 F_n(x,a)=\int_{\mathbb{C}}e^{-\frac{\bar{z}^2}{2}+\sqrt{2}\bar{z}x}\varphi_n(z)e^{-|z|^2}d\lambda(z),
 \end{equation}
 where $d\lambda$ is the Lebesgue measure on the complex plane.
 \end{proposition}
 \begin{proof}
 As a consequence of Theorem \ref{BSO} we have  \begin{equation}
F_n(x,a)=\pi^{1/4}e^{\frac{x^2}{2}}\mathcal{B}^{-1}\left(\sum_{j=0}^{n}C_j(n,a)k_{b_j}(z)\right)(x).
\end{equation} Therefore, using the expression of the Bargmann transform inverse we obtain the result.
 \end{proof}
 Now we present further observations involving the limit function $\tilde{F}(x,a)$:
 \begin{remark}
 Let $a>1$, we have, with easy calculations that 
 \begin{equation}
\displaystyle \frac{\partial}{\partial z} \mathcal{B}(\tilde{F})=\frac{ia}{\sqrt{2}} \mathcal{B}(\tilde{F}).
 \end{equation}
 Moreover, it holds that
  \begin{equation}
\displaystyle \frac{\partial^\ell}{\partial z^\ell} \mathcal{B}(\tilde{F})=\left(\frac{ia}{\sqrt{2}} \right)^\ell\mathcal{B}(\tilde{F}).
 \end{equation}
 \end{remark}

 \begin{proposition} \label{pexp}
 Let us consider the infinite operator differential operator given by
 $$\displaystyle \exp\left(\frac{\partial}{\partial z}\right):=\sum_{\ell=0}^{\infty}\frac{1}{\ell!} \left(\frac{\partial}{\partial z}\right)^\ell. $$ Then
 \begin{equation}
 \exp\left(\frac{\partial}{\partial z}\right)\mathcal{B}(\tilde{F})(z)=e^{\frac{ia}{\sqrt{2}}} \mathcal{B}(\tilde{F})(z).
 \end{equation}
 \end{proposition}
 \begin{proof}
 We observe that
 \begin{equation}
 \displaystyle \sum_{\ell=0}^{\infty}\frac{1}{\ell!}\frac{\partial^\ell}{\partial z^\ell} \mathcal{B}(\tilde{F})(z)=\mathcal{B}(\tilde{F})(z)\sum_{\ell=0}^{\infty}\frac{1}{\ell!} \left(\frac{ia}{\sqrt{2}} \right)^\ell.
 \end{equation}
 Therefore, it is clear that

 $$ \exp\left(\frac{\partial}{\partial z}\right)\mathcal{B}(\tilde{F})=e^{\frac{ia}{\sqrt{2}}} \mathcal{B}(\tilde{F}).$$

 \end{proof}
 \begin{corollary}
 For every $a>1$,
 the normalized reproducing kernel $k_{-ia/\sqrt{2}}$ is an eigenfunction of the exponential operator $\exp\left(\frac{\partial}{\partial z}\right)$ with eigenvalue $e^{\frac{ia}{\sqrt{2}}}$. In particular, we have

 \begin{equation}
 \exp\left(\frac{\partial}{\partial z}\right)k_{-ia/\sqrt{2}}=e^{\frac{ia}{\sqrt{2}}} k_{-ia/\sqrt{2}}.
 \end{equation}

 \end{corollary}
 \begin{proof}
This fact can be proved as a direct application of Proposition \ref{pexp}.
 \end{proof}

 \begin{proposition}
 Let $a>1$ and $n\geq 1$. We consider the generating function given by
 \begin{equation}
 \displaystyle \Psi_a(z,x):=\sum_{n=1}^{\infty}\frac{z^n}{n!}\tilde{F_n}(x,a);\quad z\in\mathbb{C}, \quad x\in\mathbb{R}.
 \end{equation}
 For every $z\in \mathbb{C}$ the function $\Psi_a(z,\cdot)$ belongs to $L^2(\mathbb{R})$. Moreover, we have
 \begin{equation}
 \displaystyle \left|\Psi_a(z,x)\right|\leq e^{(1+a)|z|-x^2/2}, \quad z\in \mathbb{C}, \quad x\in \mathbb{R}.
 \end{equation}
 \end{proposition}
 \begin{proof}
 We note that
 $$\displaystyle \left|\tilde{F_n}(x,a)\right|\leq (1+a)^ne^{-x^2/2}, \quad  x\in\mathbb{R.}$$
 So, it is clear that
 $$\int_\mathbb{R} \left|\tilde{F_n}(x,a)\right|^2dx \leq (1+a)^{2n}\int_\mathbb{R} e^{-x^2}dx=\sqrt{\pi}(1+a)^{2n}<+\infty.$$
 We have also  $$\displaystyle \frac{1}{n!}\left|z^n\tilde{F_n}(x,a)\right|\leq \frac{|z|^n}{n!}(1+a)^ne^{-x^2/2}, \quad  x\in\mathbb{R.}$$
 Thus, we obtain $$
 \displaystyle \left|\Psi_a(z,x)\right|\leq e^{(1+a)|z|-x^2/2}, \quad z\in \mathbb{C}, \quad x\in \mathbb{R}.
 $$
 Therefore, from the previous estimate it is clear that $\Psi_a(z,\cdot)\in L^2(\mathbb{R})$ for every $z\in\mathbb{C}$.
 \end{proof}
% \begin{remark}
 %It is possible to consider different generalizations of the super-oscillating %functions $\tilde{F_n}(x,a)$. We can consider the following extensions:
 %\begin{enumerate}
 %\item Let $\alpha>0$ and consider the weighted functions $$G_{n,\alpha}(x,a):=e^{-\alpha x^2}F_n(x,a).$$
% \item Let $(h_k(x))_{k\geq 0}$ denote the classical Hermite functions, we can consider the functions
%\begin{equation}\label{HermiteSO}
  %\mathtt{H}_{k,n}(x,a)=h_k(x)F_n(x,a),\quad k\geq 0, \quad n\geq 0.
%\end{equation}

 %\item We can also replace the classical superoscillating sequence $F_n(x,a)$ by a generalized Fourier superoscillating sequence $Y_n(x,a)$ and study their related functions:
 %$$\tilde{Y_n}(x,a):=e^{-x^2/2}Y_n(x,a).$$
 %\end{enumerate}
 %\end{remark}

\subsection{Second case: Normalized Hermite functions}
In this section we study how the Segal-Bargmann transform acts on some wave functions.
\begin{definition}
 Let $h_k(x)$ denote the normalized Hermite functions and let us set $$\mathtt{H_{k,n}}(x)=h_k(x)F_n(x,a), \quad \forall k\geq 0, n\geq 1.$$
\end{definition}

\begin{remark}
For $k=0$ we obtain the wave functions considered in the previous subsection:
$$\mathtt{H}_{0,n}(x,a)=\tilde{F}_n(x,a), \forall x\in \mathbb{R}.$$
\end{remark}
The action of Segal-Bargmann transform on the functions $\mathtt{H_{k,n}}(x)$ is given by the following result:
\begin{theorem}\label{ActionHkn}
For every $z\in\mathbb{C}$ and $k,n=0,1, \cdots$ we have
 \begin{equation}
\mathcal{B}(\mathtt{H}_{k,n})(z)=\frac{1}{\sqrt{k!}}  \displaystyle \sum_{j=0}^{n}C_j(n,a)\left(z+\frac{i}{\sqrt{2}}(1-2j/n)\right)^ke^{\frac{iz}{\sqrt{2}}(1-2j/n)-\frac{1}{4}(1-2j/n)^2}.
 \end{equation}
 \end{theorem}
 \begin{proof}
 For every fixed $k\geq 0$ and $n\geq 1$ we have
 \begin{align*}
 \displaystyle  \mathcal{B}(\mathtt{H}_{k,n})(z)& =\left(\frac{1}{\pi}\right)^{1/4}\int_\mathbb{R}e^{-\frac{1}{2}(z^2+x^2)+\sqrt{2}zx}\mathtt{H_{k,n}}(x) dx\\
&=\left(\frac{1}{\pi}\right)^{1/4}\int_\mathbb{R}e^{-\frac{1}{2}(z^2+x^2)+\sqrt{2}zx}h_k(x)F_n(x,a)dx \\
&=\sum_{j=0}^{n}C_j(n,a)\left(\left(\frac{1}{\pi}\right)^{1/4}\int_\mathbb{R} h_k(x)e^{-\frac{1}{2}(z^2+x^2)+\sqrt{2}zx} e^{i(1-2j/n)x}dx \right)\\
& \hspace{-26mm} =e^{-\frac{z^2}{2}}\sum_{j=0}^{n}C_j(n,a)e^{\frac{1}{2}\left(z+\frac{i(1-2j/n)}{\sqrt{2}}\right)^2}\left(\left(\frac{1}{\pi}\right)^{1/4}\int_\mathbb{R}h_k(x)e^{-\frac{1}{2}\left(x^2+\left( z+\frac{i(1-2j/n)^2}{\sqrt{2}}\right)^2 \right)+x\sqrt{2}(z+\frac{i(1-2j/n)}{\sqrt{2}}} dx \right).\\
 \end{align*}
 Hence, developing further the computations we obtain the following

  \begin{align*}
\displaystyle \hspace{-8mm}\mathcal{B}(\mathtt{H}_{k,n})(z)&=e^{-\frac{z^2}{2}}\sum_{j=0}^{n}C_j(n,a) e^{\frac{1}{2} \left( z+\frac{i(1-2j/n)}{\sqrt{2}}\right)^2 }\left(\frac{1}{\pi}\right)^{1/4}\int_\mathbb{R} h_k(x)e^{-\frac{1}{2}\left(x^2+\left( z+\frac{i(1-2j/n)}{\sqrt{2}}\right)^2 \right)+x\sqrt{2}(z+\frac{i(1-2j/n)}{\sqrt{2}})} dx \\
&=e^{-\frac{z^2}{2}}\sum_{j=0}^{n}C_j(n,a) e^{\frac{1}{2} \left( z+\frac{i(1-2j/n)}{\sqrt{2}}\right)^2 }\mathcal{B}[h_k]\left( z+\frac{i(1-2j/n)}{\sqrt{2}}\right)\\
 &=\frac{1}{\sqrt{k!}}  \displaystyle \sum_{j=0}^{n}C_j(n,a)\left(z+\frac{i}{\sqrt{2}}(1-2j/n)\right)^ke^{\frac{iz}{\sqrt{2}}(1-2j/n)-\frac{1}{4}(1-2j/n)^2} .\\
 \end{align*}

 \end{proof}
   \begin{corollary}\label{WeyApp}
 Let $a>1$, for every $z\in \mathbb{C}$ we denote by $e_k(z):=\frac{z^k}{\sqrt{k!}}$ the orthonormal basis of the Fock space $\mathtt{F}(\mathbb{C})$. Then, we have
  \begin{equation}
 \displaystyle \mathcal{B}(\mathtt{H}_{k,n})(z)=\sum_{j=0}^{n}C_j(n,a)k_{b_j}(z)e_k(z-b_j)=\sum_{j=0}^{n}C_j(n,a)\mathcal{W}_{b_j}{e_k}(z),
 \end{equation}
 where $b_j=-\frac{i}{\sqrt{2}}(1-2j/n)$, $k_{b_j}$ is the normalized reproducing kernel and $\mathcal{W}_{b_j}$ is the Weyl operator on the Fock space $\mathtt{F}(\mathbb{C})$.
 \end{corollary}
 \begin{proof}
For every $k\geq 0$ we use the notation $e_k(z):=\frac{z^k}{\sqrt{k!}}$ for the classical othonormal basis of the Fock space $\mathtt{F}(\mathbb{C})$. Then applying Theorem \ref{ActionHkn} we have
 \begin{align*}
 \displaystyle \mathcal{B}(\mathtt{H}_{k,n})(z)&=\frac{1}{\sqrt{k!}}  \displaystyle \sum_{j=0}^{n}C_j(n,a)\left(z+\frac{i}{\sqrt{2}}(1-2j/n)\right)^ke^{\frac{iz}{\sqrt{2}}(1-2j/n)-\frac{1}{4}(1-2j/n)^2}\\
 &=\sum_{j=0}^{n}C_j(n,a)k_{b_j}(z)e_k(z-b_j)\\
 &=\sum_{j=0}^{n}C_j(n,a)\mathcal{W}_{b_j}{e_k}(z).\\
 \end{align*}

 \end{proof}
 \begin{remark}
 For $k=0$ we recover the computations obtained for the Bargmann transform action on the superoscillating functions $\tilde{F}_{n}(x,a)$.
 \end{remark}
 In the next result, for every $k\geq 0$, we use the following notation
\begin{equation}
 \mathtt{H}_k(x):=\lim_{n\longrightarrow \infty} \mathtt{H}_{k,n}(x)=h_k(x)e^{iax}, \quad x\in \mathbb{R}.
\end{equation}
 \begin{theorem}\label{BHK}
 Let $a>1$, for every $z\in \mathbb{C}$ and $k\geq 0$. Then, we have
 \begin{equation}
\displaystyle \mathcal{B}(\mathtt{H}_k)(z)=\frac{1}{\sqrt{k!}} e^{\frac{iza}{\sqrt{2}}-\frac{a^2}{4}} \left(z+\frac{ia}{\sqrt{2}}\right)^k, \quad \forall k\geq 0.
 \end{equation}
 \end{theorem}
 \begin{proof}
 We have
 \begin{align*}
 \displaystyle \mathcal{B}(\mathtt{H}_{k})(z)&=\left(\frac{1}{\pi}\right)^{1/4}\int_\mathbb{R}e^{-\frac{1}{2}(z^2+x^2)+\sqrt{2}zx}\mathtt{H_{k}}(x) dx\\
&=\left(\frac{1}{\pi}\right)^{1/4}\int_\mathbb{R}e^{-\frac{1}{2}(z^2+x^2)+\sqrt{2}zx}h_k(x)F(x,a)dx \\
&=\left(\frac{1}{\pi}\right)^{1/4}\int_\mathbb{R} h_k(x)e^{-\frac{1}{2}(z^2+x^2)+x(\sqrt{2}z+ia)}dx \\
&=e^{-\frac{z^2}{2}-\frac{1}{2}(z+\frac{ia}{\sqrt{2}})^2} \left(\frac{1}{\pi}\right)^{1/4}\int_\mathbb{R}h_k(x)e^{-\frac{1}{2}\left(x^2+\left(z+\frac{ia}{\sqrt{2}}\right)^2 \right)+x\sqrt{2}(z+\frac{ia}{\sqrt{2}})} dx .\\
 \end{align*}
 However, we already know by classical properties of Segal-Bargmann transform that
 \begin{align*}
 \displaystyle \left(\frac{1}{\pi}\right)^{1/4}\int_\mathbb{R}h_k(x)e^{-\frac{1}{2}\left(x^2+\left(z+\frac{ia}{\sqrt{2}}\right)^2 \right)+x\sqrt{2}(z+\frac{ia}{\sqrt{2}})} dx &=\mathcal{B}(h_k)\left(z+\frac{ia}{\sqrt{2}}\right)\\
 &=\frac{1}{\sqrt{k!}}\left(z+\frac{ia}{\sqrt{2}}\right)^k.\\
 \end{align*}
 Thus, inserting these computations in the previous ones we obtain
 $$ \displaystyle \mathcal{B}(\mathtt{H}_k)(z)=\frac{1}{\sqrt{k!}} e^{\frac{iza}{\sqrt{2}}-\frac{a^2}{4}} \left(z+\frac{ia}{\sqrt{2}}\right)^k, \quad \forall k\geq 0.$$
 \end{proof}
 As a consequence of the previous result we get:
 \begin{corollary}
 It holds that
 \begin{equation}
\mathcal{B}(\mathtt{H}_k)(z)=\mathcal{W}_{- \frac{ia}{\sqrt{2}}}e_k(z), \quad \forall k\geq 0, \quad z\in \mathbb{C}.
 \end{equation}
 \end{corollary}
 \begin{proof}
 We recall that an orthonormal basis of the Fock space is given by the family of functions $e_k(z):=\frac{z^k}{\sqrt{k!}}$. Therefore, by setting $b=-\frac{ia}{\sqrt{2}}$ we have
 \begin{align*}
 \displaystyle
\mathcal{B}(\mathit{H}_k)(z)&= \frac{1}{\sqrt{k!}} e^{\frac{iza}{\sqrt{2}}-\frac{a^2}{4}} \left(z+\frac{ia}{\sqrt{2}}\right)^k\\
&=k_b(z)e_k(z-b)\\
 &=\mathcal{W}_b[e_k](z).\\
 \end{align*}
 \end{proof}
  \begin{theorem}
Let $a>1$ and $k\in \mathbb{N}$, we use the same notations as above and get
 \begin{equation}
 \lim_{n\rightarrow \infty} \sum_{j=0}^{n}C_j(n,a)\mathcal{W}_{b_j}{[e_k]}(z)=\mathcal{W}_b[e_k](z),
 \end{equation}
  where $b_j=-\frac{i}{\sqrt{2}}(1-2j/n)$ and $b=-\frac{ia}{\sqrt{2}}$.
 \end{theorem}
 \begin{proof}
 We know that the action of the Weyl operator $\mathcal{W}_b$ on functions in the Fock space remains entire. Since $(e_k)_{k\geq 0}$ is an orthonormal basis of the Fock space, then the functions $\mathcal{W}_b[e_k]$ are entire functions by construction for every $k$. Therefore, the result is obtained by applying the supershift property.
 \end{proof}
 \begin{definition}
 Let $a>1$ and $\ell=0,1,\cdots$ we introduce the functions defined by
 \begin{equation}
 \Phi_{\ell,a}(z):=\mathcal{B}\left(\frac{\mathtt{H}_\ell}{\sqrt{\ell!}}\right)(z),
 \end{equation}
 for every $z\in\mathbb{C}$. 
 \end{definition}
 Then, we have the following result:
 \begin{theorem}\label{ActPhi}
   The family of functions $(\Phi_{\ell,a})_{\ell \geq 0}$ form an Appell system with respect to the following operator $$\displaystyle T_a:=\left(\frac{\partial}{\partial z}-\frac{ia}{\sqrt{2}}\right).$$ More precisely, we have
 \begin{equation}
 T_a\Phi_{\ell,a}(z)=\Phi_{\ell-1,a}(z),
 \end{equation}
 for every $z\in\mathbb{C}$.
 \end{theorem}
 \begin{proof}
Using Theorem \ref{BHK} we have

$$\Phi_{\ell,a}(z)= \frac{1}{\ell !} \left(z+\frac{ia}{\sqrt{2}}\right)^\ell e^{\frac{iza}{\sqrt{2}}-\frac{a^2}{4}}, \quad \forall z\in\mathbb{C}.$$

Therefore, thanks to the Leibniz rule we have
\begin{align*}
\displaystyle \frac{\partial}{\partial z} \Phi_{\ell,a}(z) &= \frac{1}{\ell !}\left( \ell \left(z+\frac{ia}{\sqrt{2}}\right)^{\ell-1} e^{\frac{iza}{\sqrt{2}}-\frac{a^2}{4}}  +\frac{ia}{\sqrt{2}}\left(z+\frac{ia}{\sqrt{2}}\right)^\ell e^{\frac{iza}{\sqrt{2}}-\frac{a^2}{4}}\right)\\
&= \Phi_{\ell-1,a}(z)+\frac{ia}{\sqrt{2}}\Phi_{\ell, a}(z)\\
\end{align*}
As a consequence of these calculations we obtain

$$T_a (\Phi_{\ell,a})(z)=\Phi_{\ell-1,a}(z).$$

 \end{proof}
Now we can easily prove the following fact:
\begin{corollary}
For every $\ell\geq 0$ we have

 \begin{equation}
 T_a^\ell\Phi_{\ell,a}(z)=\Phi_{0,a}(z)= e^{\frac{iza}{\sqrt{2}}-\frac{a^2}{4}}=k_{-\frac{ia}{\sqrt{2}}}(z),
 \end{equation}
 for every $z\in\mathbb{C}$.
\end{corollary}
\begin{proof}
By applying Theorem \ref{ActPhi} we know that $T_a\Phi_{\ell,a}=\Phi_{\ell-1,a}$, for every $\ell=1,2, \cdots.$ Hence, repeating this process $\ell$-times we obtain 
$$T_a^\ell \Phi_{\ell,a}(z)=\Phi_{0,a}(z)=k_{-\frac{ia}{\sqrt{2}}}(z).$$
\end{proof}
\subsection{Third case: General $L^2$-function}
Let us fix $a>1$, $\psi\in L^2(\mathbb{R})$ and $n\geq 1$. We introduce the function \begin{equation}
\Psi_{n}(x,a)=\psi(x)F_{n}(x,a),\quad \forall x\in \mathbb{R}.
\end{equation}
Using these notations we can prove the following 
\begin{proposition}
For every $n\geq 1$, it holds that 
\begin{equation}
\displaystyle \mathcal{B}\left( \Psi_{n}\right)=\sum_{j=0}^{n} C_j(n,a)\mathcal{W}_{b_j}\left[\mathcal{B}(\psi)\right],
\end{equation}
where $b_j=$ for $j=0,\cdots, n$. 
\end{proposition}
\begin{proof}
The normalized Hermite functions $(h_k(x))_k$ form an orthonormal basis of the space $L^2(\mathbb{R})$. Therefore, we can expand the wave function $\psi$ as
$$\displaystyle \psi(x)=\sum_{k=0}^{\infty} h_k(x)\alpha_k,$$ where the complex coefficients $(\alpha_k)_{k\geq 0}$ satisfy the condition $\displaystyle \sum_{k=0}^{\infty}|\alpha_k|^2<+\infty$. Hence, the Segal-Bargmann transform of $\psi$ is given by $$\displaystyle \mathcal{B}(\psi)(z)=\sum_{n=0}^{\infty}e_k(z) \alpha_k,$$ where $e_k(z)=\frac{z^k}{\sqrt{k!}}$ for every $k=0,1,\cdots$.
Now, using the definition of the function $\Psi_n(x,a)$ we observe that it can be expressed using the functions $\mathtt{H}_{k,n}(x,a)$ considered in the previous subsection. Indeed, we have 

\begin{align*}
\displaystyle \Psi_n(x,a)&=\psi(x)F_{n}(x,a)\\
&=\sum_{k=0}^{\infty} \left(h_k(x)F_n(x,a)\right)\alpha_k\\
&=\sum_{k=0}^{\infty} \mathtt{H}_{k,n}(x,a) \alpha_k.\\
\end{align*}

Therefore, since $\mathcal{B}$ is an unitary operator from $L^2(\mathbb{R})$ onto $\mathtt{F}(\mathbb{C})$ we obtain 
$$\displaystyle \mathcal{B}\left[\Psi_n(x,a)\right](z)= \sum_{k=0}^{\infty} \mathcal{B} \left[\mathtt{H}_{k,n}(x,a)\right] \alpha_k.$$

We know by Corollary \ref{WeyApp} that the action of Segal-Bargmann on the functions $\left(\mathtt{H}_{k,n}(x,a)\right)_{k\geq 0, n\geq1}$ is given by
$$\mathcal{B}\left[\mathtt{H}_{k,n}(x,a)\right](z)=\sum_{j=0}^{n}C_j(n,a)\mathcal{W}_{b_j}[e_k](z). $$
Therefore, using the previous formula we deduce that 

\begin{align*}
\displaystyle \mathcal{B}(\Psi_n)&=\sum_{k=0}^{\infty} \mathcal{B}(\mathtt{H}_{k,n})\alpha_k\\
&=\sum_{k=0}^{\infty} \left(\sum_{j=0}^{n}C_j(n,a)\mathcal{W}_{b_j}[e_k]\right)\alpha_k\\
&=\sum_{j=0}^{n}C_j(n,a)\mathcal{W}_{b_j}\left(\sum_{k=0}^{\infty} e_k \alpha_k\right)\\
&=\sum_{j=0}^{n}C_j(n,a)\mathcal{W}_{b_j}\left(\mathcal{B}(\psi)\right).\\
\end{align*}
\end{proof}
Let $\psi\in L^2(\mathbb{R})$, we set $$\Psi(x,a)=\lim_{n\longrightarrow \infty} \psi_n(x,a):=\psi(x,a)e^{iax}.$$
\begin{proposition}
It holds that
\begin{equation}
\mathcal{B}(\Psi)(z)=\mathcal{W}_{b}(\mathcal{B}(\psi))(z),
\end{equation} 
for every $z\in \mathbb{C}$ with  $b=-\frac{ia}{\sqrt{2}}$.
\end{proposition}
\begin{proof}
The argument follows using direct calculations involving the expression of the Weyl operator.
\end{proof}
 \begin{proposition}\label{WeCon}
 Let $a>1$ and $\psi\in L^2(\mathbb{R})$. Then

 \begin{equation}
\lim_{n\rightarrow \infty} \sum_{j=0}^{n}C_j(n,a)\mathcal{W}_{b_j}{(\mathcal{B}(\psi))}(z)=\mathcal{W}_b(\mathcal{B}(\psi))(z),
 \end{equation}
  where $b_j=-\frac{i}{\sqrt{2}}(1-2j/n)$ and $b=-\frac{ia}{\sqrt{2}}$.
 \end{proposition}
 \begin{proof}
This is a direct consequence of the supershift property combined with the fact that Weyl operator produces an entire function in the Fock space. 
 \end{proof}
 \begin{proposition}
 Let $a>1$. Then, for every $f\in\mathtt{F}(\mathbb{C})$ it holds that \begin{equation}
\lim_{n\rightarrow \infty} \sum_{j=0}^{n}C_j(n,a)\mathcal{W}_{b_j}[f](z)=\mathcal{W}_b[f](z),
 \end{equation}
  where $b_j=-\frac{i}{\sqrt{2}}(1-2j/n)$ and $b=-\frac{ia}{\sqrt{2}}$.
 \end{proposition}
 \begin{proof}
 Since $f\in\mathtt{F}(\mathbb{C})$, by Segal-Bargmann characterization there exists a unique function $\psi\in L^2(\mathbb{R})$ such that we have $f=\mathcal{B}[\psi]$. Therefore, we can easily conclude using Theorem \ref{WeCon}.

 \end{proof}
 
 We can use superoscillating functions in order to approximate functions of the Fock space. For example, we represent a function in the Fock space as a pointwise limit involving Weyl operators and the coefficients of superoscillation functions:

 \begin{theorem}
 Let $a>1$. Then, for every $g\in\mathtt{F}(\mathbb{C})$, there exists a unique function $f\in \mathtt{F}(\mathbb{C})$ such that for every $z\in\mathbb{C}$ we have

 \begin{equation}
 g(z)=\lim_{n\rightarrow \infty} \sum_{j=0}^{n}C_j(n,a)\mathcal{W}_{b_j}[f](z),
 \end{equation}
  where $b_j=-\frac{i}{\sqrt{2}}(1-2j/n)$.
 \end{theorem}
 \begin{proof}
 We recall that for every $b\in\mathbb{C}$ the Weyl operator $$\mathcal{W}_b:\mathtt{F}(\mathbb{C})\longrightarrow \mathtt{F}(\mathbb{C})$$ is an isometric isomorphism such that
 $$(\mathcal{W}_{b})^{-1}=\mathcal{W}_{-b}.$$

 Therefore, since $g\in\mathtt{F}(\mathbb{C})$  there exists a unique function $f\in \mathtt{F}(\mathbb{C})$ such that we have

 $$ g(z)=\mathcal{W}_{-\frac{ia}{\sqrt{2}}}[f](z)=\lim_{n\rightarrow \infty} \sum_{j=0}^{n}C_j(n,a)\mathcal{W}_{b_j}[f](z).$$

 \end{proof}
 \begin{example}
 Let $a>1$, we have 
 \begin{equation}
 k_{-ia/\sqrt{2}}(z)=\lim_{n\rightarrow \infty} \sum_{j=0}^{n}C_j(n,a)\mathcal{W}_{b_j}[1](z)=\lim_{n\rightarrow \infty} \sum_{j=0}^{n}C_j(n,a)k_{b_j}(z).
 \end{equation}
 \end{example}
 \begin{example}
 For every $k=0,1, \cdots$ we have 
  $$ \left(z+\frac{ia}{\sqrt{2}}\right)^ke^{\frac{iaz}{\sqrt{2}}-\frac{a^2}{4}}
  =\lim_{n\rightarrow \infty} \sum_{j=0}^{n}C_j(n,a)\mathcal{W}_{b_j}[z^k].$$
In fact, it is enough to observe that
  
  $$\mathcal{W}_{-\frac{ia}{\sqrt{2}}}[z^k]:=\left(z+\frac{ia}{\sqrt{2}}\right)^ke^{\frac{iaz}{\sqrt{2}}-\frac{a^2}{4}}.$$
 \end{example}
 \section{Inverse Segal-Bargmann transform and entire superoscillation}
 \setcounter{equation}{0}
 In this section we study how the inverse Segal-Bargmann transform acts on the entire superoscillation sequence $F_n(z,a)$ in \eqref{CompSuper} written in the form

 \begin{equation}
 \displaystyle F_n(z,a)=\sum_{j=0}^{n}C_j(n,a)e^{iz(1-\frac{2j}{n})}, 
 \end{equation}
 for every $z\in\mathbb{C}$.
 A first interesting observation relating $F_n(z,a)$ to the Fock kernel $K(z,w)$ is given by the following:
 
 \begin{proposition}
 Let $a>1$ and $n\geq 1$. For every $j=0,1,\cdots, n$ we set $w_j=-i(1-\frac{2j}{n})$. Then, we have 
 \begin{equation}\label{KernelRT}
 \displaystyle F_n(z,a)=\sum_{j=0}^{n}C_j(n,a)K(z,w_j)=\sum_{j=0}^{n}C_j(n,a)K_{w_j}(z),
 \end{equation}
 for every $z\in\mathbb{C}$. Moreover, the entire superoscillation sequence $F_n(z,a)$ belongs to the Fock space $\mathtt{F}(\mathbb{C})$.
 \end{proposition}
 \begin{proof}
 The expression \eqref{KernelRT} is clear from the definitions of the entire superoscillation sequence $F_n(z,a)$ and the Fock kernels $K(z,w_j)=K_{w_j}(z)$. For the second part we note that $K_{w_j}\in \mathtt{F}(\mathbb{C})$ for every $j=0,1,\cdots, n$. Therefore, since $F_n(z,a)$ is a linear combination of $(K_{w_j})_{0\leq j\leq n}$ we conclude that the entire superoscillation is a function in the Fock space.
 \end{proof}
 \begin{remark}
 The formula \eqref{KernelRT} shows that the superoscillation function $F_n(z,a)$ can be represented as a linear combination of the Fock reproducing kernels $K(z,w_j)$ with coefficients $C_j(n,a)$. In particular, this opens some interesting questions to investigate on how superoscillations appear in the Representer Theorem used in machine learning theory, see \cite{SHS2001}.
 \end{remark}

From the previous observation it is clear that we can apply the Segal-Bargmann inverse transform to the entire superoscillation sequence $F_n(z,a)$. Next result provides explicit computations of the Bargmann inverse on $F_n(z,a)$:
\begin{theorem}\label{INTSUP}
For every $j=0,1, \cdots, n$ we set $w_j=-i(1-\frac{2j}{n})$. Then, the Bargmann inverse of $F_n(z,a)$ is given by

\begin{equation}
\mathcal{B}^{-1}(F_n(z,a))(x)=\sum_{j=0}^{n}C_j(n,a)A_{w_j}(x), 
\end{equation}
for every $x\in \mathbb{R}$. More precisely, we have 
\begin{equation}
\mathcal{B}^{-1}(F_n(z,a))(x)=e^{-\frac{x^2}{2}}\sum_{j=0}^{n}C_j(n,a)e^{\frac{(1-\frac{2j}{n})^2}{2}}e^{\sqrt{2}x i(1-\frac{2j}{n})}, \quad \forall x\in \mathbb{R}.
\end{equation}
\end{theorem}
\begin{proof}
We have 
\begin{align*}
\displaystyle F_n(z,a)&=\sum_{j=0}^{n}C_j(n,a)e^{iz(1-\frac{2j}{n})}\\
&= \sum_{j=0}^{n}C_j(n,a)K(z,w_j)\\
&=\sum_{j=0}^{n}C_j(n,a)\langle A_{w_j},A_z \rangle_{L^2(\mathbb{R})}.\\
\end{align*}

Therefore, by linearity we obtain 
\begin{align*}
\displaystyle F_n(z,a)&=\left\langle \sum_{j=0}^{n}C_j(n,a) A_{w_j},A_z \right\rangle_{L^2(\mathbb{R})} \\
&= \mathcal{B}\left(\sum_{j=0}^{n}C_j(n,a) A_{w_j}\right)(z).\\
\end{align*}
In particular, by taking the Bargmann inverse transform we get 

$$\mathcal{B}^{-1}(F_n(z,a))(x)=\sum_{j=0}^{n}C_j(n,a)A_{w_j}(x).$$
The second part of the statement follows by developing the computations using the expression of the Bargmann kernels $(A_{w_j}(x))_{0 \leq j\leq n}$.
\end{proof}
In \cite{BCSS2023} the authors studied some integral representations of the superoscillation functions. Here we discover a new integral form allowing to represent superoscillations. Our approach is built upon the use of the Segal-Bargmann transform and Bargmann kernels:
\begin{corollary}[Integral representation II]
Let $a>1$ and $n\geq 1$. We consider the linear combination of Bargmann kernels given by 
\begin{equation}
\displaystyle \sigma_n(x):=\sum_{j=0}^{n}C_j(n,a)A_{w_j}(x).
\end{equation}

Then, $\sigma_n$ belongs to $L^2(\mathbb{R})$ and we have 
\begin{equation}
\displaystyle F_n(z,a)= \left(\frac{1}{\pi}\right)^{1/4}\int_\mathbb{R}e^{-\frac{1}{2}(z^2+x^2)+\sqrt{2}zx}\sigma_n (x)dx,
\end{equation}
for every $z\in \mathbb{C}$.
\end{corollary}
\begin{proof}
It is clear from Theorem \ref{INTSUP} that 

$$F_n(z,a)=\mathcal{B}(\sigma_n(x))(z), \forall z\in \mathbb{C}.$$
Therefore, using the integral representation of the Segal-Bargmann transform we obtain the desired result.
\end{proof}
\begin{remark}
It holds that 
\begin{equation}
\displaystyle F_n(z,a)=\sum_{j=0}^{n}C_j(n,a)\langle A_{w_j}, A_z \rangle_{L^2(\mathbb{R})}, \quad \forall z\in \mathbb{C}.
\end{equation}
\end{remark}
 \section{Fourier transform and approximating sequence}
 \setcounter{equation}{0}
  In this section we compute the action of the Fourier transform $\mathcal{F}$ on the wave functions $\tilde{F}_n(x,a)$ considered in Section 3. We investigate superoscillations and approximating sequences using some results in \cite{zhu2} relating the Segal-Bargmann transform, Fourier transform, and Weyl operators.
 \subsection{Action of Fourier transform on superoscillations}

 First, we recall that the classical Fourier transform of a signal $\varphi:\mathbb{R}\longrightarrow \mathbb{C}$ is defined by
 \begin{equation}
\displaystyle  \mathcal{F}(\varphi)(\lambda):=\int_\mathbb{R} e^{-i\lambda t} \varphi(t)dt,\quad \lambda \in \mathbb{R}.
 \end{equation}
 \begin{theorem}\label{FourieFN}
 Let $a>1$, we have
 \begin{equation}
 \displaystyle \mathcal{F}(\tilde{F_n}(x,a))(\lambda)=\sqrt{2\pi}e^{-\lambda^2/2}\sum_{j=0}^{n}C_j(n,a)e^{\lambda(1-2j/n)-\frac{(1-2j/n)^2}{2} },\quad \lambda\in \mathbb{R},
 \end{equation}
 and 
 \begin{equation}\label{FF}
\displaystyle \mathcal{F}(\tilde{F}(x,a))(\lambda)=\sqrt{2\pi}e^{-\frac{(\lambda-a)^2}{2}}, \quad \lambda\in \mathbb{R}.
 \end{equation}

 \end{theorem}
 \begin{proof}
We recall that $\displaystyle \tilde{F_n}(x,a):=e^{-x^2/2}F_n(x,a),$ so applying the Fourier transform we have
\begin{align*}
\displaystyle \mathcal{F}(\tilde{F_n})(\lambda)&=\int_\mathbb{R}e^{-i\lambda x}\tilde{F_n}(x,a)dx\\
&=\int_\mathbb{R}e^{-i\lambda x}F_n(x,a)e^{-x^2/2}dx\\
&=\sum_{j=0}^{n}C_j(n,a)\int_\mathbb{R}e^{-x^2/2+ix\left[(1-2j/n)-\lambda\right]}dx.\\
\end{align*}
Hence, using the Gaussian integral given by Lemma \ref{GAUSSIANINT} with  the parameters $\alpha=1/2$ and $w=i\left[(1-2j/n)-\lambda\right]$ we obtain
\begin{align*}
\displaystyle \mathcal{F}(\tilde{F_n})(\lambda)&=\sqrt{2\pi}\sum_{j=0}^{n}C_j(n,a)e^{-\frac{\left[(1-2j/n)-\lambda\right]^2}{2}}\\
&=\sqrt{2\pi}e^{-\lambda^2/2}\sum_{j=0}^{n}C_j(n,a)e^{\lambda(1-2j/n)-\frac{(1-2j/n)^2}{2} }.\\
\end{align*}
To prove \eqref{FF} we note that $$\tilde{F}(x,a)=e^{-x^2/2}F(x,a)=e^{-x^2/2+iax},$$
 so that its Fourier transform can be computed also using Lemma \ref{GAUSSIANINT} as follows
 \begin{align*}
\displaystyle \mathcal{F}(\tilde{F})(\lambda)&=\int_\mathbb{R}e^{-i\lambda x}\tilde{F}(x,a)dx\\
&=\int_\mathbb{R}e^{-i\lambda x}F(x,a)e^{-x^2/2}dx\\
&=\int_\mathbb{R}e^{-x^2/2+ix(a-\lambda)}dx\\
&=\sqrt{2\pi}e^{-\frac{(\lambda-a)^2}{2}}.
\end{align*}
 \end{proof}
 \begin{proposition}\label{FThe}
 Let $a>1$. Then, it holds that

\begin{equation}
 \lim_{n\longrightarrow\infty} \sum_{j=0}^{n}C_j(n,a)e^{\lambda(1-2j/n)-\frac{(1-2j/n)^2}{2} }=e^{-\frac{a^2}{2}+a\lambda},\quad \lambda\in \mathbb{R}.
\end{equation}
In particular it holds that \begin{equation}\label{F1}
 \lim_{n\longrightarrow\infty} \sum_{j=0}^{n}C_j(n,a)e^{-\frac{(1-2j/n)^2}{2} }=e^{-\frac{a^2}{2}},
\end{equation}
 and
 \begin{equation}\label{F2}
 \lim_{n\longrightarrow\infty} \sum_{j=0}^{n}C_j(n,a)e^{\frac{a}{2}(1-2j/n)-\frac{(1-2j/n)^2}{2} }=1.
\end{equation}
 \end{proposition}
 \begin{proof}
The result follows directly by applying the supershift property since the function $\varphi_\lambda(a)=e^{-\frac{a^2}{2}+a\lambda}$ is entire. The two formulas \eqref{F1} and \eqref{F2} follow directly from Theorem \ref{FThe} by choosing $\lambda=0$ and $\lambda=a/2$.
 \end{proof}
 
 Following same type of computations using the normalized Hermite functions $(h_n(x))_{n\geq 0}$ we obtain the following result:
 \begin{proposition}
 Let $a>1$. Then, it holds that
  \begin{equation}
 \lim_{n\longrightarrow\infty} \sum_{j=0}^{n}C_j(n,a)T_{1-2j/n}[h_k](\lambda)=T_a[h_k](\lambda),
\end{equation}
for every $\lambda\in \mathbb{R}$.
 \end{proposition}
 \begin{proof}
 These computations are based on the fact that $$\mathcal{F}(h_k)(\lambda)=(-i)^kh_k(\lambda).$$
 \end{proof}
 
\subsection{Approximating sequences and Weyl operators}
We now apply the Segal-Bargmann transform to the approximation sequences of a function $\psi\in \mathcal{S}(\mathbb{R})$. Then, we investigate different relations with the translation operator on $L^2(\mathbb{R})$ and the Weyl operator on the Fock space. To this end, we first recall band limit and approximating sequences (see \cite{ACSSTbook2017}):
\begin{definition}[Band limit]
A function $\psi\in S(\mathbb{R})$ is said to be band limited if its Fourier transform $\mathcal{F}(\psi)$ is compactly
supported by some compact $K \subset \mathbb{R}$.
\end{definition}

\begin{definition}[Approximating sequence]
Let $a>1, n\geq 1$ and $\psi\in \mathcal{S}(\mathbb{R})$. Then, the standard approximating sequence of $\psi$ is defined to be the function given by
\begin{equation}
\displaystyle \phi_{\psi, n, a}(x):=\sum_{j=0}^{n}C_j(n,a)\psi\left(x+\left(1-2j/n\right)\right),\quad x\in \mathbb{R}.
\end{equation}
\end{definition}
\begin{lemma}[Integral representation I]
Let $a>1, n\geq 1$ and $\psi\in \mathcal{S}(\mathbb{R})$. Then, we have
\begin{equation}
\displaystyle \phi_{\psi, n, a}(x):=\frac{1}{2\pi}\int_\mathbb{R}F_n(\lambda, a)\mathcal{F}(\psi)(\lambda) e^{i\lambda x}d\lambda,\quad x\in \mathbb{R}.
\end{equation}
\end{lemma}
For a given parameter $b\in\mathbb{R}$, we use the standard notation for the classical translation operator on $L^2(\mathbb{R})$ which is defined by $T_b\varphi(x):=\varphi(x-b)$.
\begin{remark}\label{Trans}
We observe that the approximation sequence of a function $\psi\in \mathcal{S}(\mathbb{R})$ can be expressed in terms of the translation operator $T_b$ applied to $\psi$:
\begin{equation}
\displaystyle \phi_{\psi, n, a}(x):=\sum_{j=0}^{n}C_j(n,a)T_{(2j/n-1)}[\psi](x),\quad x\in \mathbb{R}.
\end{equation}

So we can prove the following result:
\begin{theorem}\label{APSTHM}
Let $\psi\in \mathcal{S}(\mathbb{R}), $ $a>1$ and $n\geq 1$. Then, we have
\begin{equation}
\mathcal{F}(\phi_{\psi, n, a})(\lambda)=\mathcal{F}(\psi)(\lambda)\left(\sum_{j=0}^{n}C_j(n,a) e^{i(1-2j/n)\lambda}\right)
\end{equation}
Moreover
\begin{equation}
\mathcal{F}(\phi_{\psi, n, a})(\lambda)=\mathcal{F}(\psi)(\lambda)F_n(\lambda,a)
\end{equation}
where $F_n(\lambda,a)$ is the classical superoscillating sequence in the frequency domain.
\end{theorem}
\begin{proof}
We use the observation in Remark \ref{Trans} and apply the linearity of Fourier in order to get

\begin{equation}
\displaystyle \mathcal{F}(\phi_{\psi, n, a})(\lambda)=\sum_{j=0}^{n}C_j(n,a)\mathcal{F}\left(T_{(2j/n-1)}[\psi]\right)(\lambda),\quad \lambda \in \mathbb{R}.
\end{equation}

Now we can use the well-known formula relating Fourier and the translation operator $T_b$ which is given by $$\mathcal{F}(T_b\psi)(\lambda)=e^{-ib\lambda}\mathcal{F}(\psi)(\lambda),\quad \lambda\in \mathbb{R}.$$
This leads to the following expression of $\mathcal{F}(\phi_{\psi, n, a})$:
\begin{align*}
\displaystyle \mathcal{F}(\phi_{\psi, n, a})(\lambda)&=\sum_{j=0}^{n}C_j(n,a)e^{i(1-2j/n)\lambda}\mathcal{F}(\psi)(\lambda) \\
&= \mathcal{F}(\psi)(\lambda)F_{n}(\lambda, a).\\
\end{align*}
\end{proof}
\begin{corollary}
Let $\psi\in \mathcal{S}(\mathbb{R}), $ $a>1$ and $n\geq 1$. Then, there exists a constant $K=K(n,a)$ depending on both $a$ and $n$ such that
\begin{equation}
||\mathcal{F}(\phi_{\psi, n, a})||_{L^2}\leq K(n,a)||\psi||_{L^2}.
\end{equation}
\end{corollary}
\begin{proof}
We apply the second part of Theorem \ref{APSTHM} and get
\begin{align*}
\displaystyle ||\mathcal{F}(\phi_{\psi, n, a})||_{L^2}^{2}&=||\mathcal{F}(\psi)(\lambda)F_n(\lambda,a)||_{L^2}^{2}\\
&\leq (1+a)^{2n} ||\mathcal{F}(\psi)||_{L^2}^2\\
&\leq (1+a)^n||\psi||_{L^2}.\\
\end{align*}
\end{proof}
\begin{theorem}
We consider the normalized Gaussian function $\psi(x)=\frac{1}{\sqrt{2\pi}}e^{-x^2/2}$. Then, it holds that:
\begin{enumerate}
\item[i)] $\displaystyle  \mathcal{F}(\phi_{\psi, n, a})(\lambda)=\tilde{F_n}(\lambda,a), \quad n\geq 1, \lambda\in \mathbb{R};$
\item[ii)] $\displaystyle\int_\mathbb{R}\phi_{\psi, n, a}(x)dx=1, \quad n\geq1 ;$
\item[iii)] $\mathcal{F}^2(\phi_{\psi,n,a})(\xi)=2\pi \phi_{\psi,n,a}(-\xi), \quad \xi\in \mathbb{R}.$
\end{enumerate}
\end{theorem}
\begin{proof}
\begin{enumerate}
\item[i)] If $\psi(x)=\frac{1}{\sqrt{2\pi}}e^{-x^2/2}$ we know that its Fourier transform is given by $$\mathcal{F}(\psi)(\lambda)=e^{-\lambda^2/2},\quad \lambda\in \mathbb{R}.$$
Then, applying the second part of Theorem \ref{APSTHM} to this particular case we obtain
\begin{align*}
\displaystyle  \mathcal{F}(\phi_{\psi, n, a})(\lambda)&=\mathcal{F}(\psi)(\lambda)F_n(\lambda,a)\\
&=e^{-\lambda^2/2}F_n(\lambda,a)\\
&=\tilde{F_n}(\lambda,a).\\
\end{align*}
\item[ii)] We can check also that the approximation sequence associated to the normalized Gaussian function $\psi(x)$ is given explicitly by
$$\displaystyle \phi_{\psi, n, a}(x)=\frac{1}{\sqrt{2\pi}}e^{-x^2/2}\sum_{j=0}^{n}C_j(n,a)e^{-x(1-2j/n)-\frac{(1-2j/n)^2}{2}},\quad \forall x\in \mathbb{R}.$$
Since $$\int_\mathbb{R} \phi_{\psi, n, a}(x)dx=\frac{1}{\sqrt{2\pi}}\sum_{j=0}^{n}C_j(n,a)e^{-\frac{(1-2j/n)^2}{2}}\int_\mathbb{R}e^{-x^2/2-x(1-2j/n)}dx,$$
developing the computations using the Gaussian integral in Lemma \ref{GAUSSIANINT} we have $$\int_\mathbb{R}e^{-x^2/2-x(1-2j/n)}dx=\sqrt{2\pi} e^{\frac{(1-2j/n)^2}{2}},\quad \forall j=0,\cdots, n$$
and this leads to the following formula
$$
\displaystyle \int_\mathbb{R} \phi_{\psi, n, a}(x)dx=\sum_{j=0}^{n}C_j(n,a)=1.
$$
\item[iii)] We apply a second time the Fourier transform to the approximation sequence $\phi_{\psi, n, a}$ and using Theorem \ref{FourieFN} we observe that
\begin{align*}
\displaystyle
\mathcal{F}^2(\phi_{\psi,n,a})(\xi)&=\mathcal{F}(\tilde{F_n}(\lambda,a))(\xi) \\
&=\sqrt{2\pi}e^{-\xi^2/2}\sum_{j=0}^{n}C_j(n,a)e^{\xi(1-2j/n)-\frac{(1-2j/n)^2}{2} }\\
&=2\pi \phi_{\psi,n,a}(-\xi).\\
\end{align*}
\end{enumerate}

\end{proof}
Now we choose as a signal function $\psi$ the Hermite functions, i.e: $\psi(x)=h_k(x)$ with $k=0,1, \cdots$. We can prove following:
\begin{proposition}
For every $k=0,1,\cdots$ and $n\geq 1$ it holds that
\begin{equation}
\mathcal{F}(\phi_{h_k,n,a})(\lambda)=(-i)^k\mathtt{H}_{k,n}(\lambda,a).
\end{equation}

\end{proposition}
\begin{proof}
 We use the well-known formula stating that Hermite functions are eigenfunction of the Fourier transform so that we have $$\mathcal{F}(h_k)=(-i)^kh_k, \quad k\geq 0.$$  Therefore, applying Theorem \ref{APSTHM} we obtain that for every $ k=0,1,\cdots$
 
 \begin{align*}
 \mathcal{F}(\phi_{h_k,n,a})(\lambda)&=(-i)^kh_k(\lambda)F_n(\lambda,a)\\
 &=(-i)^k\mathtt{H}_{k,n}(\lambda,a).\\
 \end{align*}
\end{proof}
\begin{theorem}
Let $\psi\in \mathcal{S}(\mathbb{R}), $ $a>1$ and $n\geq 1$. If we set $f=\mathcal{B}(\psi)$. Then, it holds that
\begin{equation}
\displaystyle \mathcal{B}[\phi_{\psi, n, a}](z):=\sum_{j=0}^{n}C_j(n,a){\mathcal{W}}_{\frac{2j}{n}-1}[f](z),\quad z\in \mathbb{C}.
\end{equation}
\end{theorem}
\begin{proof}
Since $\psi\in \mathcal{S}(\mathbb{R})$ it will exist a unique function $f$ in the Fock space such that $\psi=\mathcal{B}^{-1}(f)$.
Hence we obtain \begin{equation}
\displaystyle \phi_{\psi, n, a}(x):=\sum_{j=0}^{n}C_j(n,a)T_{(2j/n-1)}\mathcal{B}^{-1}(f)(x),\quad x\in \mathbb{R},
\end{equation}
By applying the Segal-Bargmann transform from both sides we get

\begin{equation}
\displaystyle \mathcal{B}[\phi_{\psi, n, a}](z):=\sum_{j=0}^{n}C_j(n,a)\left(\mathcal{B}\circ T_{(2j/n-1)}\circ \mathcal{B}^{-1}\right)[f](z),\quad x\in \mathbb{R}.
\end{equation}
We can now apply the results of Zhu in \cite{zhu2} relating the translation and Weyl operators as follows
\begin{equation}
(\mathcal{B}\circ T_b\circ \mathcal{B}^{-1})[f](z)=\mathcal{W}_b[f](z)=k_b(z)f(z-b)=e^{zb-\frac{b^2}{2}}f(z-b),\quad \forall z\in \mathbb{C}.
\end{equation}
This leads to the following
\begin{align*}
\displaystyle \mathcal{B}[\phi_{\psi, n, a}](z)&=\sum_{j=0}^{n}C_j(n,a)e^{z(2j/n-1)-\frac{(2j/n-1)^2}{2}}f(z+1-2j/n)\\
&=\sum_{j=0}^{n}C_j(n,a){\mathcal{W}}_{\frac{2j}{n}-1}[f](z)\\
\end{align*}

\end{proof}

\end{remark}

\section*{Acknowledgments}
Daniel Alpay thanks the Foster G. and Mary McGaw Professorship in Mathematical Sciences, which supported this research. Daniele C. Struppa is grateful to the Donald Bren Presidential Chair in Mathematics. Kamal Diki thanks the Grand Challenges Initiative (GCI) at Chapman University, which supported this research.

\end{document}